\documentclass[a4paper,aps,preprintnumbers,showpacs,twocolumn,superscriptaddress,nofootinbib,amsmath,amssymb]{revtex4-1}
\usepackage[dvips]{graphics}
\usepackage{color,hyperref}
\usepackage{times}
\hypersetup{
  colorlinks=true,        
  linkcolor=blue,         
  citecolor=magenta,      
  filecolor=magenta,      
  urlcolor=blue            
}

\def\beq{\begin{equation}}
\def\eeq{\end{equation}}
\def\bear{\begin{eqnarray}}
\def\ear{\end{eqnarray}}

\usepackage{enumitem}
\usepackage{graphicx,subfigure}
\usepackage{dcolumn}
\usepackage{bm}
\usepackage{color}

\begin{document}

\title{Slowly decaying resonances of massive scalar fields around Schwarzschild-de Sitter black holes}

\author{Bobir Toshmatov}
\email{bobir.toshmatov@fpf.slu.cz}
\affiliation{Institute of Physics and Research Centre of Theoretical Physics and Astrophysics, Faculty of Philosophy \& Science, Silesian University in Opava, Bezru\v{c}ovo n\'{a}m\v{e}st\'{i} 13,  CZ-74601 Opava, Czech Republic}
\affiliation{Ulugh Beg Astronomical Institute, Astronomicheskaya 33, Tashkent 100052, Uzbekistan}

\author{Zden\v{e}k Stuchl\'{i}k}
\email{zdenek.stuchlik@fpf.slu.cz}
\affiliation{Institute of Physics and Research Centre of Theoretical Physics and Astrophysics, Faculty of Philosophy \& Science, Silesian University in Opava, Bezru\v{c}ovo n\'{a}m\v{e}st\'{i} 13,  CZ-74601 Opava, Czech Republic}

\begin{abstract}
We study in special limiting cases quasinormal modes of massive scalar fields in the Schwarzschild-de Sitter black hole backgrounds. We determine the lower limit on the mass parameter of the scalar field that allows the waves with quasinormal frequencies to propagate to infinity, showing that it depends on the spacetime parameters only. Then we discuss in the large multipole number limit quasinormal modes whose frequencies can be directly related to the unstable circular photon geodesics. In the large scalar mass approximation, we demonstrate new interesting phenomenon of slowly decaying resonances, that are strongly related to the maximum of the effective potential of the massive scalar field, which is located at the static radius of the Schwarzschild-de
Sitter spacetimes, where the cosmic repulsion is just balanced by the black hole attraction.
\end{abstract}


\maketitle

\section{Introduction}

The wide variety of cosmological observations, confirmed by recent measurements of cosmic microwave background anisotropies by the space satellite observatory PLANCK, indicates that about $70\%$ of the energy content of the observable universe is represented by the so called dark energy \cite{Spe-etal:2007:ApJSuppl:,Cal-Kam:2009:NATURE:CosDarkMat,Ade-etal:2014:ASTRA:}. Restrictions coming from these observations indicate that the dark energy equation of state has to be very close to those corresponding to the vacuum energy. The repulsive vacuum energy is related to the repulsive cosmological constant, which value is thus estimated to be $\Lambda \approx 1.3\times 10^{-56}\,\mathrm{cm^{-2}}$, while the vacuum mass density $\rho_{\mathrm{vac}} \sim 10^{-29}\,\mathrm{g/cm^{3}}$ \cite{Cal-Kam:2009:NATURE:CosDarkMat}. Of course, during inflationary era, or in some phase transitions in the early Universe, the value of the cosmological constant, representing approximately the false vacuum at such an era, could be substantially higher.

The cosmological consequences of the relict cosmological constant have been studied both for the cosmological models~\cite{Mis-Tho-Whe:1973:Gra:}, and the Einstein-Straus vacuola models \cite{Stu:1983:BULAI:,Stu:1984:BULAI:,Uza-Ell-Lar:2011:GENRG2:,Gre-Lak:2010:PHYSR4:, Fle-Dup-Uza:2013:PHYSR4:,Far-etal:2015:JCAP:,Far:PDU2016}, or the more general McVittie model \cite{McV:1933:MONRAS:} of mass concentrations immersed in the expanding universe \cite{Nol:1998:PHYSR4,Nol:1999:CLAQG:,Nan-Las-Hob:2012:MONRAS:, Kal-Kle-Mar:2010:PHYSR4:,Lak-Abd:2011:PHYSR4:,Sil-Fon-Gua:2013:PHYSR4:,Nol:2014:CLAQG:}.
It has been shown that the repulsive cosmological constant can have an important role also in astrophysical processes (optical effects, accretion, jets) related to supermassive black holes in active galactic nuclei~\cite{Stu:2005:MODPLA:,Stu-Cal:1991:GENRG2:,Stu-Hle:2000:CLAQG:,Lak:2002:PHYSR4:, Bak-etal:2007:CEURJP:,Ser:2008:PHYSR4:,Mul:2008:GENRG2:FallSchBH, Sch-Zai:2008:0801.3776:CCTimeDelay,Stu-Hle:1999:PHYSR4:,Stu-Hle:2002:ActaPhysSlov:,Stu-Sla:2004:PHYSR4:,
Kra:2004:CLAQG:,Kra:2005:DARK:CCPerPrec,Kra:2007:CLAQG:Periapsis,
Stu-Sla-Hle:2000:ASTRA:,Sla-Stu:2005:CLAQG:,
Rez-Zan-Fon:2003:ASTRA:,Kag-Kun-Lam:2006:PHYLB:SolarSdS,Ali:2007:PHYSR4:EMPropKadS}. The role of the relict cosmological constant in motion of interacting galaxies has been studied in~\cite{Stu-Sch:2011:JCAP:,Stu-Sch:2012:IJMD:}, using an appropriately defined Pseudo-Newtonian potential~\cite{Stu-Kov:2008:INTJMD:}.

In all these studies an important role is played by the static radius (sometimes called also turn-around radius) where the gravitational attraction of central mass is just balanced by the cosmic repulsion, defining thus a natural boundary of gravitationally bound systems in an expanding universe. Such an idea is supported by the polytropic spherical models of galactic halos that must have their extension limited by the static radius \cite{Stu-Hle-Nov:2016:PHYSR4:}.

Here we test the role of the static radius, caused by the cosmic repulsion in the Schwarzschild-de Sitter (SdS) black hole spacetimes for the massive scalar fields, especially in the case of the quasinormal frequencies that can give interesting observational signatures of the specific features of the black hole spacetimes influenced by the cosmic repulsion. So far, several aspects of the massive (massless) scalar field in the background of the SdS and Schwarzschild-anti-de Sitter (SadS) black holes have been studied in many papers, such as~\cite{Cardoso:PRD2003,Konoplya:PRD:2002,Suneeta:PRD2003,Cardoso-Konoplya:PRD:2003, Molina:PRD2004,Choudhury:PRD2004,Konoplya:JHEP2004, Zhidenko:CQG2004,Yoshida:PRD2004,Kanti:PRD2005,Chang2006}. In the present paper we briefly study the yet uncovered properties. We discuss general properties of the effective potential of the massive scalar field in the SdS spacetimes, concentrating on its asymptotic behaviour. Namely, we give the critical value of the mass parameter of the scalar field, when the effective potential loses its potential wall part. We then study the quasinormal oscillatory modes of the massive scalar field, focusing on the special limiting case of the eikonal limit (large multipole number limit), when the quasinormal mode frequencies are directly related to the characteristics of the unstable null circular geodesics of the underlying spacetime~\cite{Cardoso:PRD2009}. Recently, it has been shown that the eikonal limit predicts long-lived quasinormal modes in the ultracompact objects~\cite{Cardoso:PRD2014}, or the so called trapping polytropes~\cite{Stuchlik:JCAP2017}, allowing for existence of trapped null geodesics. In the present paper, we point out that in the large mass approximation of the scalar fields in the SdS spacetimes the long-lived quasinormal modes can also exist, being directly related to the existence of the static radius. To study the slowly damping modes we use the standard approaches introduced and applied in the paper \cite{Hod:PLB.761:2016}.

The present paper is organized as follows: in section~\ref{sec-sds}, we shortly summarize properties of the SdS spacetimes. In section~\ref{sec-kg}, the Klein-Gordon evolutionary equation for the massive scalar fields is determined for the SdS spacetimes. Specific conditions on the mass parameter of the scalar field, related to the spacetime parameters, are given in section~\ref{sec-massive qnm}. In section~\ref{sec-qnm ups}, the quasinormal modes are determined in the eikonal limit of large values of the multipole moment, and in section~\ref{sec-massive limit}, the quasinormal resonance modes are considered in the limit of large mass parameter of the scalar field. Concluding remarks are presented in section~\ref{sec-conc}. Throughout the paper we use the geometric system of units $c=G=1$ and a spacelike signature $(-,+,+,+)$.

\section{SdS black hole spacetime}\label{sec-sds}

The line element of the spherically symmetric SdS black hole spacetime takes in the standard spherical coordinates the form
\begin{eqnarray}\label{spacetime}
ds^2=-f(r)dt^2+\frac{1}{f(r)}dr^2+r^2(d\theta^2+\sin^2\theta d\phi^2),
\end{eqnarray}
where the so called lapse function reads
\begin{eqnarray}
f(r)=1-\frac{2M}{r}-\frac{\Lambda}{3}r^2\ .
\end{eqnarray}
Here $M$ is the mass of the black hole, while $\Lambda$ is the cosmological constant. For convenience we introduce the new dimensionless parameter
\begin{eqnarray}
 y=\frac{1}{3}\Lambda M^2 ,
\end{eqnarray}
and turn into the dimensionless coordinates $t/M\rightarrow t$ and $r/M\rightarrow r$ \cite{Stu-Sla-Kov:2009:CLAQG:}.

The coordinate singularities of the SdS black hole spacetime are determined by the relation $f(r)=0$, thus, by solving this equation one can realize that when $0<y<y_{crit}\equiv 1/27$, the SdS black hole spacetime has two event horizons, located at~\cite{Stu:1983:BULAI:}
\begin{eqnarray}\label{hor}
&&r_h=\frac{2}{\sqrt{3y}}\cos\frac{\pi+\xi}{3}\ ,\\
&&r_c=\frac{2}{\sqrt{3y}}\cos\frac{\pi-\xi}{3}\ ,
\end{eqnarray}
where
\begin{eqnarray}
\xi=\arccos(3\sqrt{3y})\ .
\end{eqnarray}
At the radii $r_h$ and $r_c$, the black hole and cosmological horizons of the spacetime are located, respectively; there is $r_c>r_h$. The SdS spacetime is dynamic under the black hole event horizon ($r<r_h$) and above the cosmological event horizon ($r>r_c$). In the static region of the SdS black hole spacetimes a static radius exists at
\begin{eqnarray}
r_s=y^{-1/3} ,
\end{eqnarray}
where the gravitational attraction of the black hole is just balanced by the cosmic repulsion of the vacuum energy \cite{Stu:1983:BULAI:}.

For $y=y_{crit}$, the black hole spacetime has only one degenerate horizon where the black hole and the cosmological horizons coincide at the location of the unstable photon sphere ($r_h=r_c=r_{ph}\equiv3$) \cite{Stu-Hle:1999:PHYSR4:}.

For $y>y_{crit}$, both horizons disappear and the SdS spacetime becomes dynamic for all $r>0$, representing a naked singularity spacetime.

For the cosmologically estimated value of the relict cosmological constant, $\Lambda \approx 1.3\times 10^{-56}\,\mathrm{cm^{-2}}$, and the astrophysically estimated mass of the most supermassive black holes, going up to the value $M \sim 6\times10^{10}M_{\odot}$ \cite{Zio:2004:A}, (or even for mass parameter related to the large galaxy clusters, $M \sim 10^{16}M_{\odot}$), the location of the black hole and the cosmological horizons can be approximated with very high precession by the relations
\begin{eqnarray}
r_h=2, \qquad r_c=y^{-1/2} .
\end{eqnarray}
Therefore, in such situations $r_c\gg r_h$.

\section{Massive scalar field in the SdS black hole spacetime}\label{sec-kg}

In curved spacetimes, the scalar field with mass $\mu$ is governed by the Klein-Gordon equation
\begin{eqnarray}\label{KG}
(\Box-\mu^2)\Psi(t,r,\theta,\phi)=0\ ,
\end{eqnarray}
where
\begin{eqnarray}
 \Box\equiv\partial_\nu(\sqrt{-g}\partial^\nu)/\sqrt{-g} ,
\end{eqnarray}
and $g$ is the determinant of the spacetime metric (\ref{spacetime}). In order to separate the variables, we use the spacetime (time and axial) symmetries and assume the wave function $\Psi(t,r,\theta,\phi)$ to be  harmonically time dependent (and axisymmetric), taking the form
\begin{eqnarray}\label{WF}
\Psi(t,r,\theta,\phi)=\frac{1}{r}\psi(r)Y_\ell^m(\theta,\phi)e^{i\omega t}\ ,
\end{eqnarray}
where $Y_\ell^m(\theta,\phi)$ is the so called spherical harmonic function of degree $\ell$ and order $m$, related to the angular coordinates $\theta$ and $\phi$. Introducing the standard "tortoise" radial coordinate $x$ by the relation
\begin{eqnarray}\label{tortoise}
dx=\frac{dr}{f(r)}\ ,
\end{eqnarray}
and by using the characteristic eigenvalue of the angular eigenfunction $Y_\ell(\theta,\phi)$ given by $\ell(\ell+1)$, where the inequality $\ell\geq |m|$) has to be satisfied, we can write the Klein-Gordon equation (\ref{KG}) in the familiar wave-like form
\begin{eqnarray}\label{WE}
\left[\frac{\partial^2}{\partial x^2}+\omega^2-V(r)\right]\psi(x)=0\ ,
\end{eqnarray}
where $\omega=Re(\omega)+i Im(\omega)$. The real part of the frequency, $Re(\omega)$, represents the frequency of real oscillations, while the imaginary part, $Im(\omega)$, characterizes damping (or growing) rate. $V(r)$ is the effective potential that characterizes the dynamics of the massive scalar field in the SdS spacetime; it takes the form
\begin{eqnarray}\label{potential}
V(r)&&=f(r)\left[\frac{\ell(\ell+1)}{r^2}+\frac{f'(r)}{r}+\mu^2\right]\nonumber\\
&&=\left(1-\frac{2}{r}-yr^2\right)\left[\frac{\ell(\ell+1)}{r^2}+\frac{2}{r^3} -2y+\mu^2\right].
\end{eqnarray}
In the case of $y=0$, we recover the potential for the massive scalar field in the Schwarzschild spacetime. Recall that in the SdS spacetimes the effective potential is degenerate, being independent of the eigenvalue $m$, and that the multipole number $\ell$ increases the height of the effective potential. General properties of the effective potential of the scalar field in the SdS spacetime were studied in  papers~\cite{Cardoso:PRD2003,Molina:PRD2004,Choudhury:PRD2004,Zhidenko:CQG2004,Kanti:PRD2005,Chang2006} and we shall not repeat them. We concentrate our attention to the uncovered properties of the quasinormal modes of the massive scalar field in the SdS spacetime.

\section{Quasinormal modes of massive scalar field in the SdS spacetime}\label{sec-massive qnm}

First, we briefly review the main properties of the effective potential for the massive perturbative scalar field governed by the potential~(\ref{potential}).
\begin{figure}[t]
\begin{center}
\includegraphics[width=0.95\linewidth]{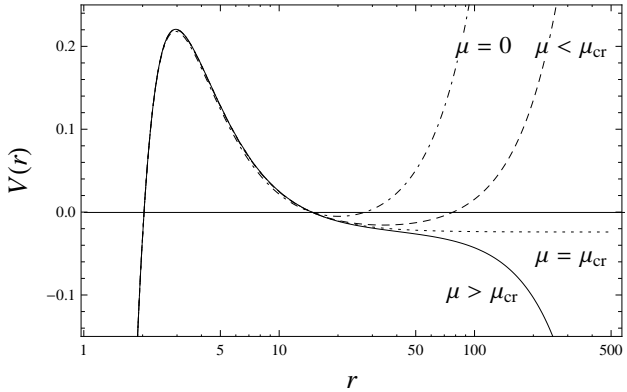}
\end{center}
\caption{\label{fig-potential} Radial profile of the effective potential of the massive scalar field in the SdS black hole spacetime for different values of the mass parameter of the field $\mu$.}
\end{figure}

It is well known that in the asymptotically flat spacetimes, the mass parameter of the scalar field cannot exceed a particular critical value \footnote{For example, for the Schwarzschild spacetimes the critical mass takes the following values: for $\ell=0$, $\ell=1$, and $\ell=2$ modes, the critical upper limit on the mass of the scalar field are $\mu_{max}\approx0.1924$, $\mu_{max}\approx0.3972$, and $\mu_{max}\approx0.6378$, respectively.}, since the effective potential tends to $\mu^2$ in large distances and, consequently, it loses the form of a barrier~\cite{Ohashi:CQG2004,Toshmatov:new}. However, in the SdS black hole spacetimes that are not asymptotically flat, the situation is different: the effective potential of the massless scalar field, or the scalar fields with relatively small mass parameter, has three zero points: two of them are located at the black hole horizon, $r_h$, and the cosmological horizon, $r_c$, and an additional zero point is at $r_e$, located above the cosmological horizon -- see Fig.~\ref{fig-potential}. Between the black hole and the cosmological horizons ($r_h$ and $r_c$), the effective potential has a barrier-like form, between the cosmological horizon $r_c$ and the zero point $r_e$ the potential is negative, and above $r_e$ the potential increases unlimitedly. The quasinormal modes of the scalar field can move between the horizons $r_h$ and $r_c$ outside the barrier given by the effective potential, however they are restricted by the increasing effective potential above the zero point $r_e$. On the other hand, if the mass parameter of the scalar field is increasing, the value of the zero point $r_e$ increases and for particular critical value of the mass parameter, $\mu_{cr}$, the third zero point $r_e$ is not defined. For $\mu>\mu_{cr}$, the effective potential is monotonically decreasing, being negative at all $r>r_c$ (see Fig.~\ref{fig-potential}). Surprisingly, this critical value of the mass depends only on the cosmological parameter
\begin{eqnarray}
\mu_{cr}=\sqrt{2y}.
\end{eqnarray}
Note that calculations have shown that even for the Reissner-Nordstr\"{o}m-de Sitter black hole spacetimes, the above given condition is satisfied and it does not depend on the value of the charge parameter of the spacetime. Thus, studying the quasinormal modes of the scalar field with mass $\mu\geq\mu_{cr}$ in the SdS black hole spacetimes is astrophysically important.

Now we give some results obtained for the quasinormal modes of the massive scalar field in the SdS spacetime. For calculations we use the sixth order WKB method given by the relation
\begin{eqnarray}\label{wkb}
\frac{i(\omega^2-V_0)}{\sqrt{-2V_0''}}+\sum_{j=2}^{6}\Lambda_j=n+\frac{1}{2}
\end{eqnarray}
where a prime ("$\prime$") denotes derivative with respect to the tortoise coordinate $x$, and $V_0$ stands for the value of the effective potential at its local maximum, $r=r_0$. The order of the WKB corrections is denoted as $j$, and $\Lambda_j$ is the correction term corresponding to the $j$th order. One can find the expressions of the $\Lambda_j$ terms in~\cite{Iyer87,Konoplya:PRD2003}.

\begin{figure*}[t]
\begin{center}
\includegraphics[width=0.47\linewidth]{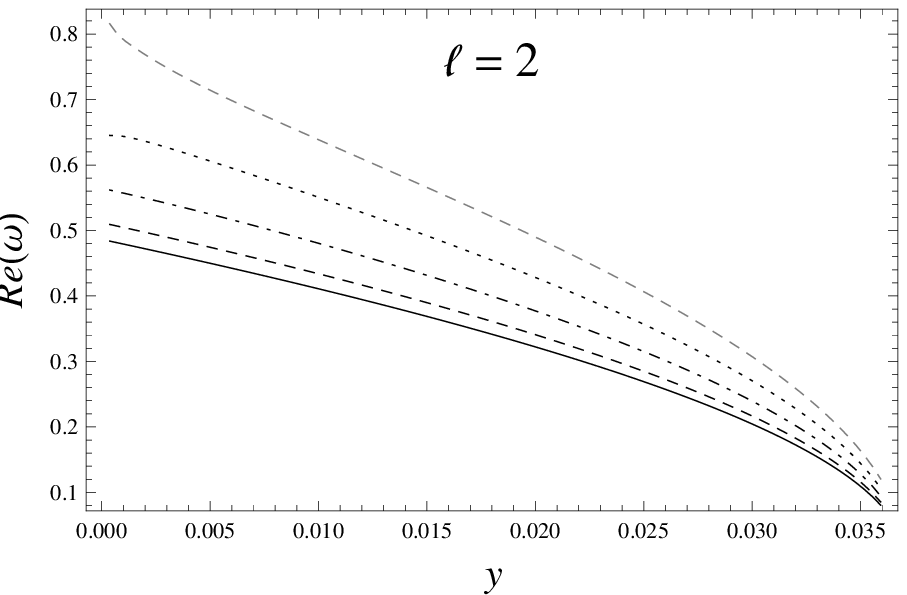}
\includegraphics[width=0.47\linewidth]{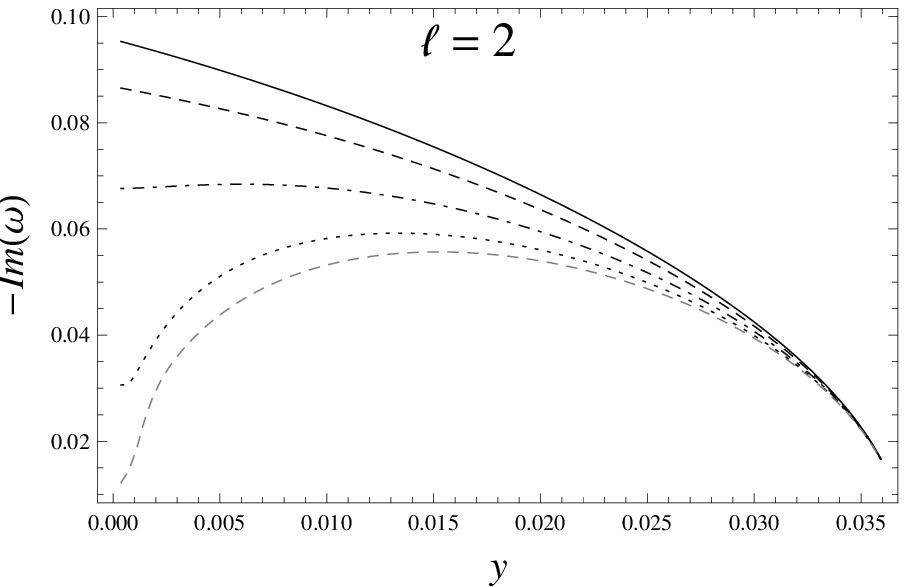}
\end{center}
\caption{\label{fig-qnm-massive} Dependence of the quasinormal frequencies of massive scalar fields in the SdS spacetimes on the cosmological parameter $y$ in the mode $\ell=2$, $n=0$ and with mass of the scalar field: $\mu=0.1$ -- black, solid, $\mu=0.3$ -- black, dashed, $\mu=0.5$ -- black, dotdashed, $\mu=0.7$ -- black, dotted, and $\mu=0.9$ -- grey, dashed.}
\end{figure*}

Since the quasinormal frequencies of the massless scalar field in the background of the SdS black holes have been studied by several authors~\cite{Cardoso:PRD2003,Molina:PRD2004,Choudhury:PRD2004,Zhidenko:CQG2004,Kanti:PRD2005,Chang2006}, we shortly present the effect of the mass of the scalar field to quasinormal frequencies in Fig.~\ref{fig-qnm-massive}.
One can see from Fig.~\ref{fig-qnm-massive} that increasing of the mass parameter of the scalar field increases the frequencies of the real oscillations and decreases their damping rate. In other words, the waves of the more massive scalar field live longer than the ones with the less massive one. With increasing value of the cosmological parameter, the real part of the quasinormal frequencies decreases, but the imaginary part decreases only for small values of the mass parameter, while it demonstrates existence of a local maximum for sufficiently large values of the mass parameter.

\section{Quasinormal modes related to unstable null circular geodesics}\label{sec-qnm ups}

In \cite{Cardoso:PRD2009} it was shown that in the eikonal limit, i.e., in the large multipole number limit, $\ell>>1$, the quasinormal frequencies can be directly studied by using the characteristics of the unstable circular null geodesics of the underlying spacetime. Then, the real part of the quasinormal frequencies is defined by the angular velocity at the unstable null circular geodesic, $\Omega_c$, while its imaginary part is determined by the Lyapunov exponent $\lambda$ which is related to the instability time scale of the near-circular null orbit. These two quantities are determined by the relations
\begin{eqnarray}\label{eikonal}
Re(\omega)+i Im(\omega)\approx\Omega_c\ell-i\left(n+\frac{1}{2}\right)|\lambda|\ ,
\end{eqnarray}
The angular velocity of the unstable circular null geodesic $\Omega_c$, and the Lyapunov exponent $\lambda$ are given by the relations
\begin{eqnarray}\label{eikonal1}
\Omega_c=\sqrt{\frac{f_c}{r_c^2}}\ , \qquad \lambda=\sqrt{\frac{f_c(f'_c-r_cf''_c)}{2r_c}}\ .
\end{eqnarray}
The radius of the unstable photon circular orbit at the static spherically symmetric spacetimes is determined by the relation
\begin{eqnarray}\label{photon orbit}
2f_c-r_cf'_c=0\ .
\end{eqnarray}
For the SdS spacetime, the radius of the unstable photon circular orbit is located at $r_{ph}=3$, as in the Schwarzschild spacetime, and it does not depend on the cosmological parameter. By evaluating for the unstable null circular geodesics the angular velocity $\Omega_c$ and the Lyapunov exponent $\lambda$~(\ref{eikonal1}) at $r_{ph}=3$, and inserting them into~(\ref{eikonal}), we obtain
\begin{eqnarray}\label{eikonal2}
\omega=\sqrt{\frac{1}{27}-y}\left[\ell- i\left(n+\frac{1}{2}\right)\right].
\end{eqnarray}
\begin{figure*}[t]
\begin{center}
\includegraphics[width=0.47\linewidth]{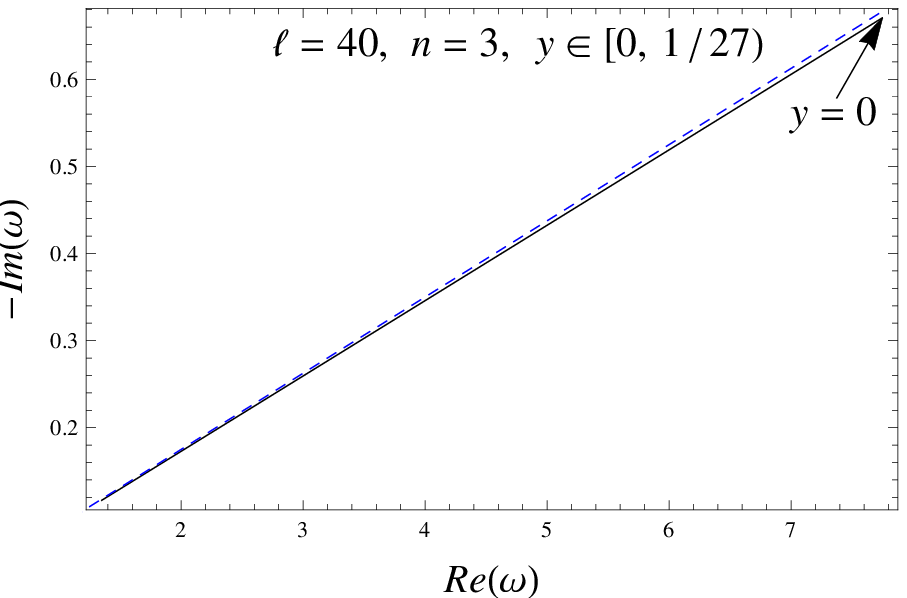}
\includegraphics[width=0.47\linewidth]{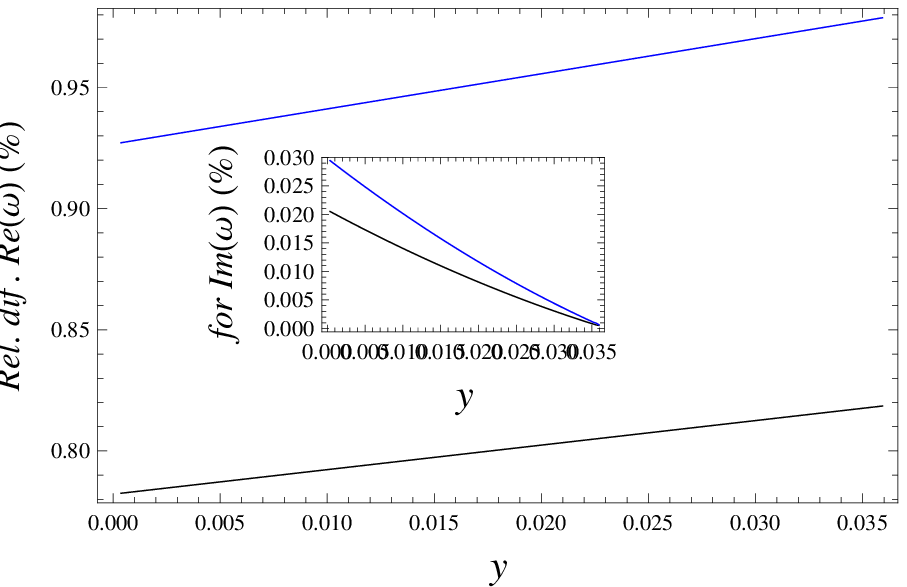}
\end{center}
\caption{\label{fig-eikonal} \textbf{Left panel:} Quasinormal frequencies of the scalar field related to the unstable null circular geodesics around SdS black hole, determined by the analytical expression~(\ref{eikonal2}) (blue, dashed) and by the numerical sixth order WKB method (black) in the large multipole number limit for $\ell=40$, $n=3$ mode. \textbf{Right panel:} Relative difference (Rel. dif.) in the quasinormal frequencies given by the analytical expression~(\ref{eikonal2}) and the sixth order WKB method for $\ell=50$, $n=3$ (blue) and $\ell=60$, $n=3$ (black) modes.}
\end{figure*}
One can see from Fig.~\ref{fig-eikonal} that this expression has very good agreement with numerical results in the large multipole number limit. With increasing value of the multipole number $\ell$, accuracy of the eikonal expression~(\ref{eikonal2}) increases.  For $y=0$, we recover the expressions for the Schwarzschild black hole case. Moreover, in the SdS black hole backgrounds, the scalar field oscillates with smaller quasinormal frequencies than in the pure Schwarzschild backgrounds. One can see from~(\ref{eikonal2}) that when the cosmological parameter is equal to the critical value, $y=y_c\equiv1/27$, i.e, when the black hole and the cosmological horizons coincide at the unstable photon circular orbit~\cite{Stu-Hle:1999:PHYSR4:}, the quasinormal frequencies vanish.


\section{The quasinormal resonances of massive scalar fields in the large mass regime}\label{sec-massive limit}

In this section we study the quasinormal resonance frequencies of the massive scalar field in the SdS black hole background in the eikonal large mass approximation~\cite{Hod:PLB.761:2016,Hod:CQG2015}
\begin{eqnarray}\label{large-mass}
\mu^2\gg\ell^2\ .
\end{eqnarray}
In this limit, the effective potential (\ref{potential}) can be approximated as
\begin{eqnarray}\label{potential-limit}
V(r)=\left(1-\frac{2}{r}-yr^2\right)\mu^2+O\left[\left(\frac{1}{M\mu}\right)^2\right].
\end{eqnarray}
The maximum of this effective potential is located at the static radius
\begin{eqnarray}\label{maximum}
r_0=r_s\equiv y^{-1/3}
\end{eqnarray}
that impressive astrophysical significance~\cite{Stu-Sch:2011:JCAP:,Stuchlik-Hledik:PRD:2016}.
\begin{figure*}[t]
\begin{center}
\includegraphics[width=0.47\linewidth]{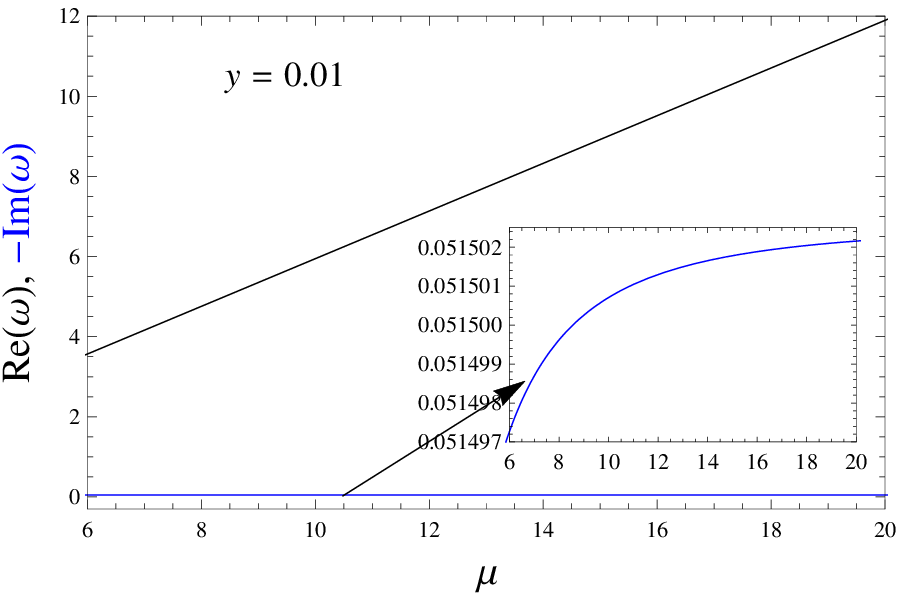}
\includegraphics[width=0.47\linewidth]{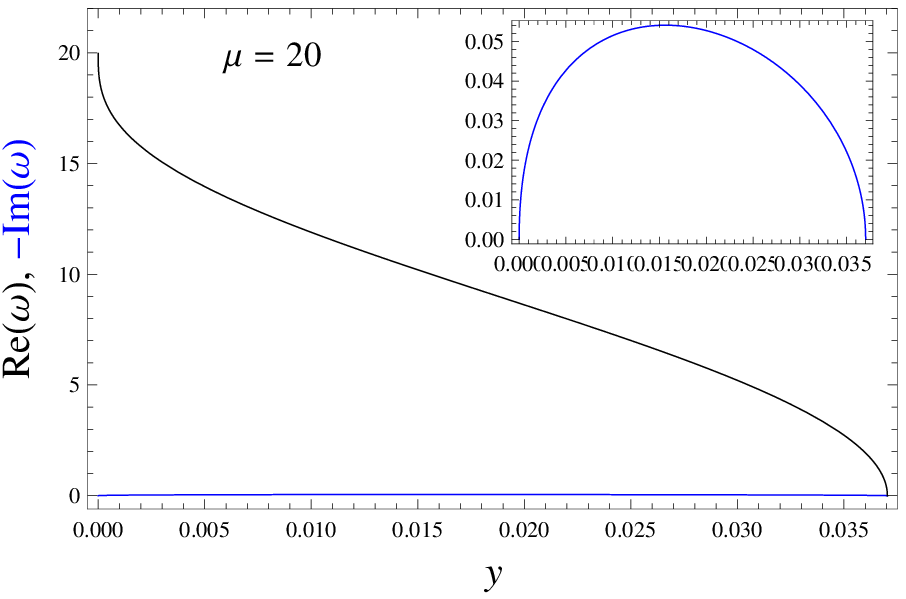}
\end{center}
\caption{\label{fig-01} Fundamental quasinormal resonant frequencies of the massive scalar field in the SdS black hole back ground treated in the large mass regime.}
\end{figure*}
By using the first order WKB method~(\ref{wkb}), we can find the quasinormal frequencies of the massive scalar field in the SdS spacetime in the large mass regime with the effective potential given by relation (\ref{potential-limit}). By substituting (\ref{potential-limit}) and (\ref{maximum}) into (\ref{wkb}), we obtain the expression for the characteristic resonance frequencies in the form
\begin{eqnarray}\label{resonances}
\frac{\omega^2-\left(1-3y^{1/3}\right)\mu^2}{2\sqrt{3y} (1-3y^{1/3})\mu}=-i\left(n+\frac{1}{2}\right).
\end{eqnarray}
By solving Eq.~(\ref{resonances}) analytically, one can find that the real and imaginary parts of the quasinormal resonant frequency can be given in the form
\begin{eqnarray}
&&Re(\omega)=\frac{\mu\sqrt{1-3y^{1/3}}\sqrt{1+\sqrt{1+\frac{12y}{\mu^2} \left(n+\frac{1}{2}\right)^2}}}{\sqrt{2}}\ ,\label{freq-1}\\
&&Im(\omega)=-\frac{\mu\sqrt{1-3y^{1/3}}\sqrt{-1+\sqrt{1+\frac{12y}{\mu^2} \left(n+\frac{1}{2}\right)^2}}}{\sqrt{2}}.\nonumber\\\label{freq-2}
\end{eqnarray}
One can see from Eqs~(\ref{freq-1}) and~(\ref{freq-2}) that these expressions are valid only in the black hole spacetimes ($y<y_{crit}\equiv1/27$). Eqs.~(\ref{freq-1}),~(\ref{freq-2}) and Fig.~\ref{fig-01} show that the real part of the quasinormal resonant frequency, $Re(\omega)$, is an (almost linearly) monotonically increasing function of the dimensionless mass parameter of the scalar field $\mu$. Moreover, the imaginary part of the quasinormal resonant frequency, $Im(\omega)$, is also increasing function of $\mu$, but after some intermediate value of $\mu$ it stops increasing at a value that is almost negligible relative to the real part of the quasinormal frequency. Therefore, we can deduce that these are clearly slowly decaying resonances.

In the Schwarzschild limit, $y\rightarrow0$, we arrive to
\begin{eqnarray}\label{sch-limit}
Re(\omega)\rightarrow\mu\ , \qquad Im(\omega)\rightarrow0\ .
\end{eqnarray}
In the limit of $y\rightarrow y_{crit}\equiv1/27$, both the real and imaginary parts of the quasinormal resonant frequency vanish
\begin{eqnarray}\label{crit-limit}
Re(\omega)\rightarrow0\ , \qquad Im(\omega)\rightarrow0\ .
\end{eqnarray}
The characteristic relaxation time of the system is defined by the imaginary part of the fundamental quasinormal frequency ($\tau_{relax}=1/Im(\omega),n=0$). Thus the characteristic relaxation time of the quasinormal resonances in the SdS black hole spacetimes reads
\begin{eqnarray}\label{limit-2}
\tau_{relax}=\frac{\sqrt{2}}{\mu\sqrt{1-3y^{1/3}}\sqrt{-1+\sqrt{1+\frac{3y}{\mu^2}}}}.
\end{eqnarray}
\begin{figure}[t]
\begin{center}
\includegraphics[width=0.90\linewidth]{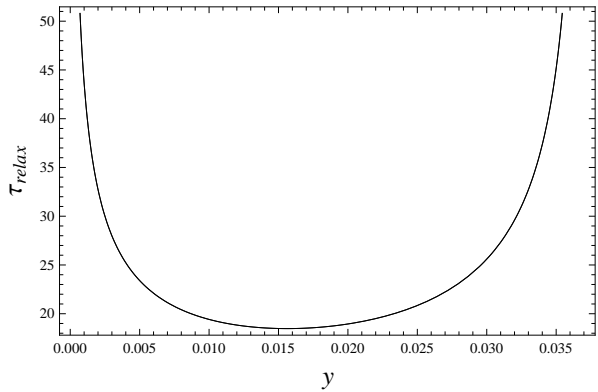}
\end{center}
\caption{\label{fig-02} Dependence of the characteristic relaxation time $\tau_{relax}$ of the quasinormal resonances of massive scalar field in the field of the SdS black hole on the cosmological parameter $y$.}
\end{figure}
One can see from (\ref{sch-limit}),~(\ref{crit-limit}) and~(\ref{limit-2}), or Fig.~\ref{fig-02}, that the characteristic relaxation time scale diverges when $y=0$ and $y=y_{crit}$. However, as we mentioned already in~(\ref{crit-limit}), in the extremal SdS black hole backgrounds ($y=y_{crit}$), the real part of the quasinormal frequencies vanishes too. Thus, slowly damping quasinormal resonances of the massive scalar fields in the SdS black hole backgrounds in the large mass regime exist for small values of the cosmological parameter $y$ and are related to the static radius of the background giving thus additional relevance of the notion of static radius.

\section{Conclusion}\label{sec-conc}

In the present paper we have studied the dynamics of the massive or massless scalar fields in the SdS spacetimes. First, we have given the lower limit on the value of the mass of the scalar field for the waves with quasinormal frequencies related to the effective potential barrier between the black hole and the cosmological horizons of the SdS spacetime, and enabling the wave to reach observer at infinity without meeting any other potential wall. Surprisingly, the critical mass of the scalar field determining the lover limit depends only on the dimensionless cosmological parameter given by the relation $y=\Lambda M^2/3$. Moreover, the same statement is relevant to the Reissner-Nordstr\"{o}m-de Sitter spacetimes.

Second, we have determined the quasinormal frequencies of the oscillations in the eikonal approximation related to the unstable photon circular orbits and shown that with increasing cosmological parameter $y$ of the SdS spacetime, both frequency of the oscillations of the quasinormal modes and the damping rate decrease.

Third, we have studied the large mass approximation of the effective potential of the massive scalar field that is governed by the spacetime function $f(r)$, similarly to the eikonal limit governed by the circular null geodesic structure. We have found an interesting and relevant new phenomenon of the quasinormal resonance frequencies of the massive scalar fields in the large mass regime, that are shown to be directly related to the static radius of the black hole spacetimes. We have explicitly demonstrated existence of slowly decaying quasinormal resonances in the SdS spacetimes with astrophysically relevant very small values of the cosmological parameter $y$ that are similar to the long-lived quasinormal perturbations in ultracompact objects. For the Schwarzschild black holes these quasinormal modes correspond to pure real oscillations with frequency equal to the mass of the scalar field, $Re(\omega)=\mu$. The characteristic relaxation time of the quasinormal resonances tends to infinity in the Schwarzschild limit, $y\rightarrow0$. In the SdS spacetimes, the relaxation time of the resonance modes can give signature of the presence of the static radius and its distance from the black hole, giving thus an additional information on the extension of the gravitationally binding region~\cite{Stuchlik-Hledik:PRD:2016}.

\begin{acknowledgments}
The authors thank the Albert Einstein Centre for Gravitation and Astrophysics supported by the Czech Science Foundation under the grant No.~14-37086G, and the internal student grant of Silesian University~SGS/14/2016. B.T. is partially supported by Grant No.~VA-FA-F-2-008 of the Uzbekistan Agency for Science and Technology.
\end{acknowledgments}

\label{lastpage}

\end{document}